# Local Utility Elicitation in GAI Models

**Darius Braziunas**
Department of Computer Science
University of Toronto
Toronto, ON M5S 3H5
darius@cs.toronto.edu

**Craig Boutilier**
Department of Computer Science
University of Toronto
Toronto, ON M5S 3H5
cebly@cs.toronto.edu

## Abstract

Structured utility models are essential for the effective representation and elicitation of complex multiattribute utility functions. Generalized additive independence (GAI) models provide an attractive structural model of user preferences, offering a balanced tradeoff between simplicity and applicability. While representation and inference with such models is reasonably well understood, elicitation of the parameters of such models has been studied less from a practical perspective. We propose a procedure to elicit GAI model parameters using only "local" utility queries rather than "global" queries over full outcomes. Our local queries take full advantage of GAI structure and provide a sound framework for extending the elicitation procedure to settings where the uncertainty over utility parameters is represented probabilistically. We describe experiments using a myopic value-of-information approach to elicitation in a large GAI model.

## 1 Introduction

The increased interest in automated decision support tools in recent years has brought the problem of *automated preference elicitation* to the forefront of research in decision analysis [7, 15, 12] and AI [5, 6, 2]. Generally speaking, the goal of automated preference elicitation is to devise algorithmic approaches that will guide a user through an appropriate sequence of queries or interactions and determine enough about her preferences to make a good or optimal decision. Many models have been proposed, including those that treat responses to queries as constraints on utilities (including methods in conjoint analysis [13]) and those that use priors over utility parameters.

Crucial to preference elicitation in complex domains is the existence of utility function *structure* [11, 8]. Structure in the form of additive, multilinear, generalized additive or other models [11, 8, 1, 3] can be used to represent utility models very concisely. While additive models are by far the most commonly used in practice, *generalized additive independence (GAI) models* [8, 1, 10] have drawn more attention recently because of their additional flexibility. Unfortunately, effective elicitation procedures for GAI models have attracted far less attention than additive models. Thus, for example, recent procedures for eliciting parameters of GAI models often ignore the semantic foundations of direct queries [4]. Gonzales and Perny [10] recently addressed this problem. Using the semantic foundations of Fishburn [8] they discuss a graphical model which can be used to guide elicitation in GAI models.

In this paper, we continue the exploration of elicitation of GAI utility model parameters. One difficulty with the procedure of Gonzales and Perny [10] is its reliance on standard gamble queries involving full outcomes. In large, multiattribute domains, it can be cognitively unmanageable for a user to compare full outcomes involving more than a handful of attributes; furthermore, this fails to take advantage of the independence structure in the queries themselves. We propose a new elicitation technique that allows the parameters of a GAI model to be determined using (almost exclusively) "local" queries over a small number of attributes, while respecting the Fishburn semantics.

Our second contribution is a procedure for *partial elicitation* of utility parameters. Generally speaking, good (or even optimal) decisions can be realized without complete utility information. Rather than asking for the direct assessment of utility parameters using standard gambles as in [10], we consider simpler binary *comparison queries* over gambles. Following [6, 2], we suppose some prior over the parameters of a GAI model, and use myopic expected value of information (EVOI) to determine appropriate queries. The advantages of GAI models become very clear in such a setting, since the implied decomposition allows us to effectively compute EVOI in very large models. We demonstrate our procedure on a large (26 variable) constraint-based configuration problem, showing that it is fast enough to support interactive elicitation.

## 2 GAI Models

We begin with some standard concepts from multiattribute utility theory [11, 8].

### 2.1 Background and Notation

Assume a set of attributes $X_1, X_2, \ldots, X_n$, each with finite domains (for ease of notation we use $X_i$ to refer to its domain as well). These define a set of *outcomes* (or alternatives or consequences) $\mathbf{X} = X_1 \times \cdots \times X_n$ over which a decision maker (DM) has preferences. A preference relation $\succeq$ is a total preorder over the set of outcomes, with $\mathbf{x} \succeq \mathbf{x}'$ meaning that $\mathbf{x}$ is at least as preferred as $\mathbf{x}'$. Strict preference $\succ$ and indifference $\sim$ are defined in the usual way. Given an index set $I \subseteq \{1, \ldots, n\}$, we define $\mathbf{X}_I = \times_{i \in I} X_i$ to be the set of *partial outcomes* restricted to attributes in $I$, and $\mathbf{x}_I$ to be the same restriction of a specific outcome $\mathbf{x}$. $I^C$ denotes $I$'s complement.

User preferences can be expressed by a bounded, real-valued *utility function* $u : \mathbf{X} \mapsto \mathbb{R}$ such that $u(\mathbf{x}) \geq u(\mathbf{x}')$ iff $\mathbf{x} \succeq \mathbf{x}'$. A utility function serves as a quantitative representation of strength of preferences, and can be used to represent preferences over *lotteries* (distributions over outcomes); specifically, one lottery is preferred to another iff its expected utility is greater [14]. Let $\langle p_1, \mathbf{x}_1; \ldots; p_k, \mathbf{x}_k \rangle$ denote a lottery over $k$ outcomes with $\mathbf{x}_i$ realized with probability $p_i$ (and $\sum_i p_i = 1$). Since utility functions corresponding to $\succeq$ are unique only up to positive affine transformation, it is customary to set the utility of the best outcome $\mathbf{x}^\top$ to 1, and the utility of the worst outcome $\mathbf{x}^\perp$ to 0. Thus, if a DM is indifferent between $\mathbf{x}$ and the *standard gamble* $\langle p, \mathbf{x}^\top; 1-p, \mathbf{x}^\perp \rangle$, then $u(\mathbf{x}) = p$.

### 2.2 Additive Independence

Since the number of outcomes is exponential in the number of attributes, specifying the utility of each outcome is infeasible in most practical applications. However, $u$ can be expressed concisely if it exhibits sufficient structure. *Additive independence* [11] is one structural assumption commonly used in practice. Under a strong independence assumption—specifically, that the DM is indifferent among lotteries that have same marginals on each attribute—$u$ can be written as a sum of single-attribute *subutility functions*:

$$u(\mathbf{x}) = \sum_{i=1}^{n} u_i(x_i) = \sum_{i=1}^{n} \lambda_i v_i(x_i).$$

This simple factorization exploits subutility functions $u_i(x_i) = \lambda_i v_i(x_i)$, which themselves depend on *local value functions* $v_i$ and scaling constants $\lambda_i$. The assumed utility independence among attributes allows elicitation to proceed *locally*: specifically, the $v_i$ can be elicited independently of other attribute values. Since each attribute is utility independent, each attribute's best and worst levels can be determined separately. Formally, $x_i^\top \in X_i$ is $X_i$'s *best attribute level* if and only if

$$(x_i^\top, \mathbf{y}) \succeq (x_i^k, \mathbf{y}) \; \forall x_i^k \in X_i, \mathbf{y} \in \mathbf{X}_{i^C}.$$

The worst level $x_i^\perp$ is defined similarly. A *local preference* between $x_i^k$ and a *local gamble* $\langle p, x_i^\top; 1-p, x_i^\perp \rangle$ is well-defined since utility independence implies that $(x_i^k, \mathbf{y}) \succeq \langle p, (x_i^\top, \mathbf{y}); 1-p, (x_i^\perp, \mathbf{y}) \rangle$ for *some* $\mathbf{y} \in \mathbf{X}_{i^C}$ iff this holds for *all* such $\mathbf{y}$. Indifference for a specific $p$ implies that

$$u(x_i^k, \mathbf{y}) = p\, u(x_i^\top, \mathbf{y}) + (1-p)\, u(x_i^\perp, \mathbf{y}),$$

and therefore, because of the additive form $u$,

$$v_i(x_i^k) = p\, v_i(x_i^\top) + (1-p)\, v_i(x_i^\perp).$$

If we set $v_i(x_i^\top) = 1$, $v_i(x_i^\perp) = 0$, then $v_i(x_i^k) = p$. Local value functions $v_i(\cdot)$ can be therefore elicited using only *local* standard gamble queries that involve two local "anchor" outcomes $x_i^\top$ and $x_i^\perp$.

After performing local elicitation, we know each attribute's local value relative to the utilities of the respective anchor outcomes. What remains is to bring all the local value scales to the common global utility scale. To achieve global consistency, queries involving full outcomes are unavoidable. Essentially, we need to find the true utility of all "anchor" outcomes $x_i^\top$ and $x_i^\perp$, with respect to some default outcome $\mathbf{x}^0$. It is customary to choose the worst outcome as default outcome, and set its utility to 0. Then, eliciting

$$u_i^\top \equiv u_i(x^\top) = u(x_i^\top, \mathbf{x}_{i^C}^0),$$
$$u_i^\perp \equiv u_i(x^\perp) = u(x_i^\perp, \mathbf{x}_{i^C}^0) = 0$$

for all attributes ensures consistent scaling of subutility functions. Scaling factors $\lambda_i$, which reflect attribute contributions to the overall utility function, are simply $u_i^\top$.

### 2.3 Generalized Additive Independence

GAI models [8, 1] provide an additive decomposition of utility functions in situations where single attributes are not additively independent, but (possibly nondisjoint) subsets of attributes are. The form of a GAI model is as follows. Assume a collection $\{I_1, \ldots, I_m\}$ of (possibly intersecting) index sets such that $\cup_i I_i = \{1, \ldots, n\}$ and *local subutility functions* $u_i$ over $\mathbf{X}_{I_i}$. Then $u$ is decomposed as:

$$u(\mathbf{x}) = u_1(\mathbf{x}_{I_1}) + \ldots + u_m(\mathbf{x}_{I_m}).$$

If, say, $I_1 = \{1, 2\}$, and $I_2 = \{2, 3\}$ in a three-attribute domain, then $u(x_1, x_2, x_3) = u_1(x_1, x_2) + u_2(x_2, x_3)$.

We discuss the foundations of GAI models below, but first illustrate difficulties with generalizing local elicitation of the type suitable for additive models to GAI models [10]. In the additive case, $u_i(x_i^1) > u_i(x_i^2)$ implies that outcomes with $i$th attribute set to $x_i^1$ are preferred to outcomes with $x_i^2$, as long as the rest of attributes are kept constant. However, in GAI models we cannot draw such straightforward conclusions. Let's take our example $u(x_1, x_2, x_3) = u_1(x_1, x_2) + u_2(x_2, x_3)$. If we know that $u_1(x_1^1, x_2^1) = 10$ and $u_1(x_1^1, x_2^2) = 5$, does it imply $(x_1^1, x_2^1) \succeq (x_1^1, x_2^2)$, ceteris paribus? It turns out that because of interdependence of subutility factors, we can rewrite the utility function as follows ($f(x_2)$ is an arbitrary real-valued function):

$$\begin{aligned} &u(x_1, x_2, x_3) \\ &\quad = [u_1(x_1, x_2) + f(x_2)] + [u_2(x_2, x_3) - f(x_2)] \\ &\quad = u'_1(x_1, x_2) + u'_2(x_2, x_3). \end{aligned}$$

If $f(x_2^1) = -5$, and $f(x_2^2) = 5$, then $u'_1(x_1^1, x_2^1) = 5$ and $u'_1(x_1^1, x_2^2) = 10$, the exact opposite of $u_1(\cdot)$. Since the utility can "flow" from one subutility factor to the next through the shared attributes, the subutility values do not have an independent semantic meaning. This example illustrates that the same utility function can be decomposed in an infinite number of non-trivial ways.

The conditions under which a GAI model provides an accurate representation of a utility function were defined by Fishburn [8, 9], who introduced the model.[1] Let $\mathcal{P}$ be the set of all gambles (probability distributions) on $\mathbf{X}$, and $\mathcal{P}_I$ be the set of all gambles on $\mathbf{X}_I$. For $P \in \mathcal{P}$, $P_I$ is the marginal gamble of $P$ over $\mathbf{X}_I$. Let $\{I_1, \ldots, I_m\}$ be a collection of nonempty subsets of $\{1, \ldots, n\}$.

**Defn. 1.** The sets of attributes indexed by $I_1, \ldots, I_m$ are *(generalized) additively independent* if and only if

$$[(P_{I_1}, \ldots, P_{I_m}) = (Q_{I_1}, \ldots, Q_{I_m})] \implies P \sim Q,$$

that is, if and only if the decision maker is indifferent between two lotteries whenever their marginal distributions on $\mathbf{X}_{I_1}, \ldots, \mathbf{X}_{I_m}$ are the same.

**Theorem 1.** *[8] The GAI condition holds iff there are real-valued subutility functions $u_1, \ldots, u_m$ on $\mathbf{X}_{I_1}, \ldots, \mathbf{X}_{I_m}$ such that*

$$u(\mathbf{x}) = u_1(\mathbf{x}_{I_1}) + \ldots + u_m(\mathbf{x}_{I_m}). \tag{1}$$

The following important result relies on the notion of a *default outcome*, denoted by $\mathbf{x}^0 = (x_1^0, x_2^0, \ldots, x_n^0)$ (where each $x_i$ is set to an arbitrary value). For any $\mathbf{x}$, let $\mathbf{x}[I]$ be the outcome where attributes not in $I$ are set to the default value, but other attributes remain as in $\mathbf{x}$ (i.e., $X_i = x_i$ if $i \in I$, and $X_i = x_i^0$ if $i \notin I$). For example, if $\mathbf{x} = (x_1, x_2)$, then $\mathbf{x}[\{1\}] = (x_1, x_2^0)$.

**Theorem 2.** *[8] If GAI holds, then for all $\mathbf{x} \in \mathbf{X}$:*

$$u(\mathbf{x}) = \sum_{j=1}^{m} (-1)^{j+1} \sum_{1 \le i_1 < i_2 < \cdots < i_j \le m} u\left(\mathbf{x}\left[\bigcap_{s=1}^{j} I_{i_s}\right]\right). \tag{2}$$

This theorem captures all dependencies intrinsic to GAI utility functions. In our running example,
2[ex] $u(x_1, x_2, x_3) = u(x_1, x_2, x_3^0) + u(x_1^0, x_2, x_3) - u(x_1^0, x_2, x_3^0)$.

[1] Fishburn used the term *interdependent value additivity*; Bacchus and Grove [1] dubbed the same concept GAI, which seems to be more commonly used in the AI literature currently.

Given three arbitrary attribute sets $I_1, I_2, I_3$, we have:

$$\begin{aligned} u(\mathbf{x}) &= u(\mathbf{x}[I_1]) + u(\mathbf{x}[I_2]) + u(\mathbf{x}[I_3]) \\ &\quad - u(\mathbf{x}[I_1 \cap I_2]) - u(\mathbf{x}[I_1 \cap I_3]) - u(\mathbf{x}[I_2 \cap I_3]) \\ &\quad + u(\mathbf{x}([I_1 \cap I_2 \cap I_3]). \end{aligned}$$

As we can see, under GAI conditions, Theorem 2 provides a way to write the utility of any outcome $\mathbf{x}$ as a sum of utilities of certain other *key* outcomes. These outcomes are related to $\mathbf{x}$ in a specific way: in each of them, some attributes are set to the same levels as in outcome $\mathbf{x}$, while remaining attributes are at their default values.

Theorem 2 allows one to construct the subutility functions required in Eq. 1. If we group the addends on the right side of Eq. 2 appropriately, we can define $u_1, \ldots, u_m$ such that $u(\mathbf{x}) = \sum_{j=1}^m u_j(\mathbf{x}_{I_j})$. There is, however, more than one way to define these subutility functions. Let $\mathbf{x}_j$ denote $\mathbf{x}_{I_j}$ (the restriction of $\mathbf{x}$ to attributes in $I_j$). Fishburn [8] proposed the following construction for subutility functions:

$$\begin{aligned} u_1(\mathbf{x}_1) &= u(\mathbf{x}[I_1]), \\ u_j(\mathbf{x}_j) &= u(\mathbf{x}[I_j]) + \sum_{k=1}^{j-1} (-1)^k \sum_{1 \le i_1 < \cdots < i_k < j} u(\mathbf{x}[\bigcap_{s=1}^{k} I_{i_s} \cap I_j]). \end{aligned} \tag{3}$$

We call this the *canonical* subutility decomposition. In our trivial example, the canonical decomposition would be:

$$\begin{aligned} u_1(x_1, x_2) &= u(x_1, x_2, x_3^0) \equiv u(\mathbf{x}[I_1]); \\ u_2(x_2, x_3) &= u(x_1^0, x_2, x_3) - u(x_1^0, x_2, x_3^0) \\ &= u(\mathbf{x}[I_2]) - u(\mathbf{x}[I_1 \cap I_2]). \end{aligned}$$

In the case of three overlapping attribute sets $I_1, I_2, I_3$:,

$$\begin{aligned} &u_1(\mathbf{x}_1) = u(\mathbf{x}[I_1]), \\ &u_2(\mathbf{x}_2) = u(\mathbf{x}[I_2]) - u(\mathbf{x}[I_1 \cap I_2]), \\ &u_3(\mathbf{x}_3) = u(\mathbf{x}[I_3]) - (u(\mathbf{x}[I_1 \cap I_3]) + u(\mathbf{x}[I_2 \cap I_3])) \\ &\quad + u(\mathbf{x}[I_1 \cap I_2 \cap I_3]). \end{aligned}$$

Recall that $u(\cdot)$ denotes utility of full outcomes, whereas $u_i(\cdot)$ is defined over attributes indexed by $I_i$.

## 3 GAI Elicitation with Local Queries

If we could easily elicit utilities of key outcomes, the elicitation task would be straightforward: the utility of any $\mathbf{x}$ can be calculated using the utilities of related key outcomes via Eq. 2. This simplifies elicitation because the decision maker only has to specify utilities of key outcomes (see [10] for a relevant elicitation algorithm). Unfortunately, even key outcomes are "full" outcomes over all attributes; it is unrealistic to expect a user to assess tradeoffs involving full outcomes in domains with more than a few attributes. Therefore, just as in the elicitation of additive utility functions, we would like to separate the elicitation process into local elicitation and global scaling.

### 3.1 Local Elicitation

Assume that for each subset $I_i$ we have chosen two different "top" and "bottom" *anchor* outcomes $\mathbf{x}[I_i]^\top = (\mathbf{x}_{I_i}^\top, \mathbf{x}_{I_i^C}^0)$ and $\mathbf{x}[I_i]^\perp = (\mathbf{x}_{I_i}^\perp, \mathbf{x}_{I_i^C}^0)$.[2] In these outcomes, the attributes indexed by the set $I_i$ are set to their "top" and "bottom" levels, respectively, while the other attributes are set to the default level. We will assume that $\mathbf{x}[I_i]^\top$ is the best possible outcome and $\mathbf{x}[I_i]^\perp$ is the worst possible outcome given that attributes not in $I_i$ are set to the default level; however, in general, this does not have to be the case as long as top and bottom anchor outcomes are different.

Assuming these two anchor levels for each subset $I_i$, we can express the utility of certain outcomes in terms of anchor outcome utilities *in a local way*. First, we define $M_j$ to be the union of all the subsets that have variable $j$:

$$M_j = \bigcup_{i:j\in I_i} I_i.$$

We can think of $M_j$ as the *neighbor set* of the attribute $j$; it includes all the attributes that share subsets with the attribute $j$. Then, the *conditioning set* $C_i$ of the set $I_i$ is just the union of the neighbor sets of the attributes in $I_i$ minus the attributes in $I_i$:

$$C_i = \bigcup_{j\in I_i} M_j - I_i.$$

For example, let $u(x_1,\ldots,x_7) = u_1(x_1,x_2,x_3,x_6) + u_2(x_1,x_2,x_7)+u_3(x_2,x_4)+u_4(x_4,x_5)+u_5(x_5,x_6)$ (see Fig. 1). Then, the neighbor set of $x_4$ is $M_4 = \{2,5\}$ and the conditioning set for $I_4$ is $C_4 = \{2,6\}$.

After appropriate rearrangement of indices, an outcome $\mathbf{x}$ can be written as $(\mathbf{x}_i, \mathbf{x}_{C_i}, \mathbf{y})$, where $\mathbf{y}$ are the attributes that are neither in $I_i$ nor $C_i$. If the attributes in the conditioning set are at default level, then we have the following:

**Theorem 3.** *Under GAI conditions, if*

$$(\mathbf{x}_i, \mathbf{x}_{C_i}^0, \mathbf{y}) \sim \langle p, (\mathbf{x}_i^\top, \mathbf{x}_{C_i}^0, \mathbf{y}); 1-p, (\mathbf{x}_i^\perp, \mathbf{x}_{C_i}^0, \mathbf{y})\rangle, \text{ then}$$
$$(\mathbf{x}_i, \mathbf{x}_{C_i}^0, \mathbf{y}') \sim \langle p, (\mathbf{x}_i^\top, \mathbf{x}_{C_i}^0, \mathbf{y}'); 1-p, (\mathbf{x}_i^\perp, \mathbf{x}_{C_i}^0, \mathbf{y}')\rangle,$$

*for any* $\mathbf{y}'$. *Therefore,*

$$(\mathbf{x}_i, \mathbf{x}_{C_i}^0) \sim \langle p, (\mathbf{x}_i^\top, \mathbf{x}_{C_i}^0); 1-p, (\mathbf{x}_i^\perp, \mathbf{x}_{C_i}^0)\rangle.$$

That is, as long as attributes in the conditioning set of $I_i$ are fixed, the remaining attributes do not influence the strength of preference of local outcomes $\mathbf{x}_i$. Thus, we can perform *local* elicitation with respect to local anchors $\mathbf{x}_i^\top$ and $\mathbf{x}_i^\perp$, without specifying the levels of the $\mathbf{y}$ attributes.

**Proof** Given collection of subsets $\{I_1,\ldots,I_m\}$, let $\mathcal{C}^i$ be a partition of this collection such that $\mathcal{C}^i$ contains all $I_j$ that share some attribute with $I_i$: $\mathcal{C}^i = \{I_j : I_j \cap I_i \neq \emptyset\}$. All subsets in $\mathcal{C}^i$ contain only variables in $I_i$ and $C_i$. Thus, if

$$(\mathbf{x}_i, \mathbf{x}_{C_i}^0, \mathbf{y}) \sim \langle p, (\mathbf{x}_i^\top, \mathbf{x}_{C_i}^0, \mathbf{y}); 1-p, (\mathbf{x}_i^\perp, \mathbf{x}_{C_i}^0, \mathbf{y})\rangle,$$

then

$$\begin{aligned}
&u(\mathbf{x}_i, \mathbf{x}_{C_i}^0, \mathbf{y}) = pu(\mathbf{x}_i^\top, \mathbf{x}_{C_i}^0, \mathbf{y}) + (1-p)u(\mathbf{x}_i^\perp, \mathbf{x}_{C_i}^0, \mathbf{y}) \\
&\Longrightarrow \\
&= \sum_{I_j\in\mathcal{C}^i} u_j(\mathbf{x}_j[I_i]) + \sum_{I_j\notin\mathcal{C}^i} u_j(\mathbf{y}_j) = \sum_{I_j\notin\mathcal{C}^i} u_j(\mathbf{y}_j) + \\
&= \left[p \sum_{I_j\in\mathcal{C}^i} u_j(\mathbf{x}_j^\top[I_i]) + (1-p)\sum_{I_j\in\mathcal{C}^i} u_j(\mathbf{x}_j^\perp[I_i])\right] \\
&\Longrightarrow \\
&\sum_{I_j\in\mathcal{C}^i} u_j(\mathbf{x}_j[I_i]) = \\
&p \sum_{I_j\in\mathcal{C}^i} u_j(\mathbf{x}_j^\top[I_i]) + (1-p)\sum_{I_j\in\mathcal{C}^i} u_j(\mathbf{x}_j^\perp[I_i]) \Longrightarrow \\
&(\mathbf{x}_i, \mathbf{x}_{C_i}^0) \sim \langle p, (\mathbf{x}_i^\top, \mathbf{x}_{C_i}^0); 1-p, (\mathbf{x}_i^\perp, \mathbf{x}_{C_i}^0)\rangle. \quad \square
\end{aligned}$$

Thus, the utility of any suboutcome $\mathbf{x}_i$ of factor $i$ can be expressed locally in terms of the two anchor levels, given that attributes in the conditioning set of $i$ are set to their default values. We can now define a *local value function* $v_i(\cdot)$ such that $v_i(\mathbf{x}_i^\top) = 1$, $v_i(\mathbf{x}_i^\perp) = 0$, and $v_i(\mathbf{x}_i) = p$ iff

$$(\mathbf{x}_i, \mathbf{x}_{C_i}^0) \sim \langle p, (\mathbf{x}_i^\top, \mathbf{x}_{C_i}^0); 1-p, (\mathbf{x}_i^\perp, \mathbf{x}_{C_i}^0)\rangle.$$

We can calibrate the relative values of $v_i(\mathbf{x}_i)$ within any subutility factor (conditional on $C_i$ at default levels) using only queries over attributes in $I_i$ and $C_i$.[3] This stands in contrast to the elicitation procedure of [10] which uses full outcomes. After local elicitation, we know the conditional local values $v_i(\cdot)$ for all settings of attributes in $I_i$.

### 3.2 Global scaling

Suppose we have elicited the local value functions $v_i$ and the utilities of anchor outcomes $\mathbf{x}[I_i]^\top$ and $\mathbf{x}[I_i]^\perp$ (recall that anchor outcomes are full outcomes). Let $u_i^\top = u(\mathbf{x}[I_i]^\top)$ and $u_i^\perp = u(\mathbf{x}[I_i]^\perp)$. The utility of an arbitrary outcome $\mathbf{x}$ can now be calculated from the utilities of anchor outcomes and local value functions. By the definition of local value functions (assuming $v_i(\mathbf{x}_i) = p$),

$$(\mathbf{x}_i, \mathbf{x}_{C_i}^0, \mathbf{y}^0) \sim \langle p, (\mathbf{x}_i^\top, \mathbf{x}_{C_i}^0, \mathbf{y}^0);\ 1-p, (\mathbf{x}_i^\perp, \mathbf{x}_{C_i}^0, \mathbf{y}^0)\rangle,$$
$$(\mathbf{x}_i, \mathbf{x}_{C_i}^0, \mathbf{y}^0) \sim \langle v_i(\mathbf{x}_i),\ \mathbf{x}[I_i]^\top;\ 1-v_i(\mathbf{x}_i),\ \mathbf{x}[I_i]^\perp\rangle.$$

Therefore, for any $J_i \subseteq I_i$,

$$\begin{aligned}
u(\mathbf{x}[J_i]) &= v_i(\mathbf{x}_i[J_i])\ u_i^\top + (1 - v_i(\mathbf{x}_i[J_i]))\ u_i^\perp \\
&= (u_i^\top - u_i^\perp)\ v_i(\mathbf{x}_i[J_i]) + u_i^\perp.
\end{aligned}$$

[2] It is important to keep in mind that anchor levels are defined for each subutility factor, not individual attributes.

[3] It is important to distinguish local value functions (which are only locally calibrated) from the GAI subutility functions $u_i$, even though both are defined over the same factors.

Finally, we define the subutility functions $u_1, \ldots, u_m$ in terms of anchor outcome utilities and local value functions. Using the canonical definition (Eq. 3), we get

$$\begin{aligned} u_1(\mathbf{x}_1) &= (u_1^\top - u_1^\perp)\, v_1(\mathbf{x}_1) + u_1^\perp, \qquad (4)\\ u_j(\mathbf{x}_j) &= (u_j^\top - u_j^\perp) \cdot \\ &\left[ v_j(\mathbf{x}_j) + \sum_{k=1}^{j-1} (-1)^k \sum_{1 \le i_1 < \cdots < i_k < j} v_j(\mathbf{x}_j[\bigcap_{s=1}^{k} I_{i_s} \cap I_j]) \right] \\ &+ \left[ u_j^\perp + \sum_{k=1}^{j-1} (-1)^k \sum_{1 \le i_1 < \cdots < i_k < j} u_j^\perp \right]. \end{aligned}$$

In our small example, this gives:

$$\begin{aligned} u_1(x_1, x_2) &= (u_1^\top - u_1^\perp)\, v_1(x_1, x_2) + u_1^\perp, \\ u_2(x_2, x_3) &= (u_2^\top - u_2^\perp)\, [v_2(x_2, x_3) - v_2(x_2, x_3^0)]. \end{aligned}$$

### 3.3 Graphical Elicitation Procedure

In practice, we expect GAI models to exhibit considerable structure, and intersections between subutility factors to involve only a few variables. We propose a complete utility elicitation procedure that takes advantage of such structure. For now, we assume that a decision maker is capable of answering *direct* local standard gamble utility queries, such as "for what probability $p$ would you be indifferent between suboutcome $\mathbf{x}_i$ and a (local) standard lottery $\langle p, \mathbf{x}_i^\top; 1-p, \mathbf{x}_i^\perp \rangle$, assuming that attributes in the conditioning set $C_i$ are at default levels?" Later, we will consider more realistic local *comparison* queries.

Assume a decomposition of attributes into GAI subsets $I_1, \ldots, I_m$, and fix an order over these subsets (the order does not affect efficiency of our algorithm). We construct a directed graph whose nodes correspond to the sets $I_i$ and directed edges $(I_i, I_j)$ whenever $I_i \cap I_j \neq \emptyset$ and $i > j$.[4] Edge $(I_i, I_j)$ is labeled by $I_i \cap I_j$. Fig. 1 shows an example of a GAI graph.

After local elicitation, we have local value functions $v_i(\cdot)$. Utilities of anchor levels $u_1^\top, u_1^\perp, \ldots, u_m^\top, u_m^\perp$ have to be obtained by global queries. However, we only need to ask $2m$ such queries involving full outcomes.[5] Interestingly, this is the *same number of global queries* required for global scaling in the linearly additive case (considering each attribute as a factor in the additive case).

The general formula for defining canonical subutility functions is provided by Eq. 4. However, we can simplify it considerably due to the graphical structure of GAI attribute sets. A utility function $u'$ is strategically equivalent to $u$ if $u'$ is a positive affine transformation of $u$. Notice that the expression $u_j^\perp + \sum_{k=1}^{j-1} (-1)^k \sum u_j^\perp$ on the last line in Eq. 4 does not depend on the particular configuration $\mathbf{x}_j$. Therefore, it is simply a constant and can be eliminated. Furthermore, when $\bigcap_{s=1}^{k} I_{i_s} \cap I_j = \emptyset$, we have $v_j(\mathbf{x}_j[\bigcap_{s=1}^{k} I_{i_s} \cap I_j]) = v_j(\mathbf{x}_j[\emptyset]) = v_j(\mathbf{x}_j^0)$, i.e., the local utility of the default suboutcome. This utility is fixed at $(u(\mathbf{x}^0) - u_j^\perp)/(u_j^\top - u_j^\perp)$, and does not depend on the argument $\mathbf{x}_j$; therefore, every $v_j(\mathbf{x}_j[\bigcap_{s=1}^{k} I_{i_s} \cap I_j])$ for which $\bigcap_{s=1}^{k} I_{i_s} \cap I_j = \emptyset$ can be eliminated from Eq. 4.

[4] An undirected version of this graph is a *GAI network* [10].

[5] Only $m$ queries are required if $\mathbf{x}^0 = \mathbf{x}^\perp$.

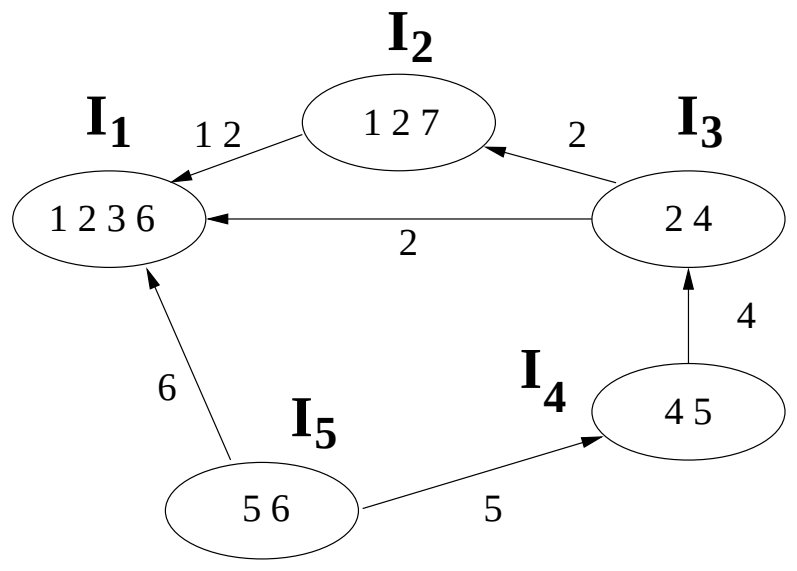

Figure 1: An Example of a GAI Graph.

Any utility function $u$ can now be rewritten as a strategically equivalent utility function $u'$:

$$u'(\mathbf{x}) = \sum_{j=1}^{m} u'_j(\mathbf{x}_j) = \sum_{j=1}^{m} \bar{v}_j(\mathbf{x}_j)(u_j^\top - u_j^\perp), \qquad (5)$$

where

$$\bar{v}_j(\mathbf{x}_j) = v_j(\mathbf{x}_j) + \sum_{k=1}^{j-1} (-1)^k \sum_{1 \le i_1 < \cdots < i_k < j} v_j(\mathbf{x}_j[\bigcap_{s=1}^{k} I_{i_s} \cap I_j]),$$

and $v_j(\mathbf{x}_j[\bigcap_{s=1}^{k} I_{i_s} \cap I_j])) = 0$, if $\bigcap_{s=1}^{k} I_{i_s} \cap I_j = \emptyset$.

To compute a (unnormalized) subutility function $\bar{v}_j(\mathbf{x}_j)$, we have to know which local suboutcomes $\mathbf{x}'_j$ are involved (in the form $\mathbf{x}_j[\bigcap_{s=1}^{k} I_{i_s} \cap I_j]$) on the right side of the equation; this amounts to finding all nonempty sets $\bigcap_{s=1}^{k} I_{i_s} \cap I_j$ and recording the corresponding sign of the local value functions in Eq. 5. The structure of subutility functions depends only on the GAI subset decomposition. Therefore, given a GAI graph, we can use the following search procedure to compute the relevant subsets needed to solve Eq. 5. We only need to do this once for each subutility factor:

**Input:** GAI attribute sets $I_1, \ldots, I_m$.
**Output:** For each subutility factor $j$, a collection of sets $L_j$, and a sign function $z_j : L_j \mapsto \{+1, -1\}$.

- For each subutility factor $j$:
- Start at node $j$ and perform a graph search along the directed arcs. The search depth is finite, so any search algorithm (e.g., breadth-first or depth-first) could be used. Set $L_j = \emptyset$.
- While $I_i \cap I_j \neq \emptyset$ (we're at node $i$)
  - let $K = \{$nodes on path from $j$ to $i\}$;
  - add $\cap_{k \in K} I_k$ to $L_j$;
  - set $z_j(\cap_{k \in K} I_k) = 1$, if depth even,
    $z_j(\cap_{k \in K} I_k)$ = -1, if depth odd.

Because of the graphical structure of GAI models, Eq. 5 now reduces to

$$\bar{v}_j(\mathbf{x}_j) = \sum_{J \in L_j} z_j(J)\ v_j(\mathbf{x}_j[J]).$$

Consider the example GAI graph in Figure 1. To compute $L_5$, we search for all non-empty intersections of the set $I_5$ with other sets. The only such sets are $I_5$ itself (at depth 0), $I_4$ (depth 1), and $I_1$ (depth 1). Therefore, $L_5 = \{\{5,6\},\{5\},\{6\}\}$, and $\bar{v}_5(x_5,x_6) = v_5(x_5,x_6) - v_5(x_5,x_6^0) - v_5(x_5^0,x_6)$. Finally, $u_5(x_5,x_6) = \bar{v}_5(x_5,x_6)(u_5^\top - u_5^\perp)$.

# 4 Elicitation under Uncertainty

We now consider *partial elicitation* of utility parameters. Generally speaking, good (or even optimal) decisions can be realized without complete utility information. Rather than asking for the direct assessment of utility parameters using standard gambles as in [10], we consider simpler binary *comparison queries* over local gambles. Following [6, 2], we suppose some prior over the parameters of a GAI model, and use myopic expected value of information (EVOI) to determine appropriate queries.

If a utility function $u$ is completely unstructured, and a prior density $\pi$ over the utility function parameters is available, the best outcome with respect to the prior is simply $\mathbf{x}^* = \arg\max_{\mathbf{x}} E^\pi[u(\mathbf{x})]$. However, we can query a user about her utility function, update the prior based on the response, and compute a new expected best outcome. If a sequence of queries can be asked, finding the best elicitation policy is a sequential decision process, providing an optimal tradeoff between query costs (the burden of elicitation) and potentially better decisions due to additional information [2]. However, such a policy is very difficult to compute, so here we adopt a myopic approach to choosing the next query [6].

We consider queries of the form "Is utility of outcome $\mathbf{x}$ greater than $l$?", denoted $\langle \mathbf{x}^q, l\rangle$; these require only $yes, no$ responses.[6] The *expected posterior utility (EPU)* of query $\langle \mathbf{x}^q, l\rangle$ is

$$\begin{aligned} EPU(\mathbf{x}^q,l) = &\Pr(yes|\mathbf{x}^q,l) \max_{\mathbf{x}} E^{yes|\mathbf{x}^q,l}\ [u(\mathbf{x})] + \\ &\Pr(no|\mathbf{x}^q,l) \max_{\mathbf{x}} E^{no|\mathbf{x}^q,l}\ [u(\mathbf{x})], \end{aligned}$$

where $\Pr(yes|\mathbf{x}^q,l)$ is the probability of response $yes$ w.r.t. the current density and $E^{yes|\mathbf{x}^q,l}$ is expectation w.r.t. the updated density given a $yes$ response. The *expected value of information* of query $\langle \mathbf{x}^q, l\rangle$ is:

$$EVOI(\mathbf{x}^q,l) = EPU(\mathbf{x}^q,l) - E[u(\mathbf{x}^*)].$$

Therefore, the best myopic query is

$$\langle \mathbf{x}^q, l\rangle^* = \arg\max_{\mathbf{x}^q} \arg\max_{l} EPU(\mathbf{x}^q,l).$$

This requires optimization over all outcomes in $\mathbf{X}$, as well as continuous optimization of the *query point* $l \in [0,1]$.

## 4.1 GAI Structure and Local Queries

GAI models allow us take advantage of the additive utility decomposition to compute EVOI. We assume that anchor utilities $u_1^\top, u_1^\perp, \ldots, u_m^\top, u_m^\perp$ are known, but the local value functions $v_1,\ldots,v_m$ are specified imprecisely via independent priors over local value function parameters. Thus, for each suboutcome $\mathbf{x}_i$ (apart from three special configurations $\mathbf{x}_i^\top, \mathbf{x}_i^\perp, \mathbf{x}_i^0$ whose local values are fixed) we have an independent prior density over possible values of $v_i(\mathbf{x}_i)$.[7] The expected value of outcome $\mathbf{x}$ is then

$$E[u(\mathbf{x})] = \sum_{j=1}^m E[u_j(\mathbf{x}_j)] = \sum_{j=1}^m (u_j^\top - u_j^\perp) E[\bar{v}_j(\mathbf{x}_j)],$$

where $E[\bar{v}_j(\mathbf{x}_j)]$ equals

$$E[v_j(\mathbf{x}_j)] + \sum_{k=1}^{j-1}(-1)^k \sum_{1\le i_1<\cdots<i_k<j} E[v_j(\mathbf{x}_j[\bigcap_{s=1}^k I_{i_s} \cap I_j])].$$

With priors over local utility functions, an appropriate form of query is "Is local utility of suboutcome $\mathbf{x}_i$ greater than $l$?", denoted as $\langle \mathbf{x}_i^q, l\rangle$. Such a query is a *local* query, because it asks a user to focus on preferences over a (usually small) subset of attributes; the values of remaining attributes do not have to be considered. Indeed, this corresponds to a binary comparison query over local outcomes and gambles, which a user can more easily assess: "do you prefer $\mathbf{x}_i$ or $\langle p, \mathbf{x}_i^\top; 1-p, \mathbf{x}_i^\perp\rangle$, assuming that attributes in the conditioning set $C_i$ are at default levels?" The best local myopic query is then

$$\begin{aligned} \langle \mathbf{x}_i^q, l\rangle^* = \arg\max_{\mathbf{x}_i^q} \arg\max_{l} \\ &\Pr(yes|\mathbf{x}_i^q,l) \max_{\mathbf{x}} E^{yes|\mathbf{x}_i^q,l}\ [u(\mathbf{x})] + \\ &\Pr(no|\mathbf{x}_i^q,l) \max_{\mathbf{x}} E^{no|\mathbf{x}_i^q,l}\ [u(\mathbf{x})]. \end{aligned}$$

We can simplify part of the equation as follows:

$$\begin{aligned} &\max_{\mathbf{x}} E^{yes|\mathbf{x}_i^q,l}\ [u(\mathbf{x})] \\ &= \max_{\mathbf{x}_i} \left[ E^{yes|\mathbf{x}_i^q,l}\ [u_i(\mathbf{x}_i)] + \max_{\mathbf{x}\text{ restr. to }\mathbf{x}_i} \sum_{j\neq i} E[u_j(\mathbf{x}_j)] \right] \\ &= \max_{\mathbf{x}_i} \left[ (u_i^\top - u_i^\perp) E^{yes|\mathbf{x}_i^q,l}\ [\bar{v}_i(\mathbf{x}_i)] + r(\mathbf{x}_i) \right], \end{aligned}$$

where $r(\mathbf{x}_i) = \max_{\mathbf{x}\text{ restr. to }\mathbf{x}_i} \sum_{j\neq i} E[u_j(\mathbf{x}_j)]$ could be computed by, say, variable elimination.

We need some additional notation. Let *dep*$(\mathbf{x}'_j)$ be the set of all suboutcomes $\mathbf{x}_j$ such that $\mathbf{x}'_j$ appears (in the form $\mathbf{x}_j[\bigcap_{s=1}^k I_{i_s} \cap I_j]$) on the right side of an expression for $\bar{v}_j(\mathbf{x}_j)$ in Eq. 5.[8] Intuitively, subutility values of outcomes in *dep*$(\mathbf{x}'_j)$ "depend" on the local utility of outcome $\mathbf{x}'_j$. Also, let $s(\mathbf{x}_j, \mathbf{x}'_j)$ be the signed unit (i.e., +1 or -1) in front of $v_j(\mathbf{x}'_j)$ on the right side of an equation for $\bar{v}_j(\mathbf{x}_j)$.

[6] The range of a utility function is assumed to be $[0,1]$.

[7] Recall $v_i(\mathbf{x}_i^\top) = 1, v_i(\mathbf{x}_i^\perp) = 0, and v_i(\mathbf{x}_i^0) = \frac{u(\mathbf{x}^0) - u_i^\perp}{u_i^\top - u_i^\perp}$.

[8] The set *dep*$(\mathbf{x}'_j)$ can be computed easily from the set $L_j$ obtained by the GAI graph search procedure.

If $\mathbf{x}_i \notin$ *dep*$(\mathbf{x}^q_i)$, then a query involving $\mathbf{x}^q_i$ will not change the expected value of $u(\mathbf{x}_i)$: $E^{yes|\mathbf{x}^q_i,l}\ [u_i(\mathbf{x}_i)] = E[u_i(\mathbf{x}_i)]$. If $\mathbf{x}_i \in$ *dep*$(\mathbf{x}^q_i)$, then the expected posterior value of $u_i(\mathbf{x}_i)$ changes only because of the change in the posterior expectation of $v_i(\mathbf{x}^q_i)$. Therefore, in such a case,

$$\begin{aligned} &E^{yes|\mathbf{x}^q_i,l}\ [\bar{v}_i(\mathbf{x}_i)] \\ &= E[\bar{v}_i(\mathbf{x}_i)] - s(\mathbf{x}_i, \mathbf{x}^q_i) E[v_i(\mathbf{x}^q_i)] + s(\mathbf{x}_i, \mathbf{x}^q_i)\ E^{yes|\mathbf{x}^q_i,l}\ [v_i(\mathbf{x}^q_i)] \\ &= s(\mathbf{x}_i, \mathbf{x}^q_i)\ E^{yes|\mathbf{x}^q_i,l}\ [v_i(\mathbf{x}^q_i)] + E[\bar{v}_i(\mathbf{x}_i)] - s(\mathbf{x}_i, \mathbf{x}^q_i)\ E[v_i(\mathbf{x}^q_i)]. \end{aligned}$$

Thus,

$$\begin{aligned} &\max_{\mathbf{x}} E^{yes|\mathbf{x}^q_i,l}\ [u(\mathbf{x})] \\ &\quad = \max_{\mathbf{x}_i} \left[ E^{yes|\mathbf{x}^q_i,l}\ [u_i(\mathbf{x}_i)] + r(\mathbf{x}_i) \right] \\ &\quad = \max \begin{cases} \max_{\mathbf{x}_i \notin dep(\mathbf{x}^q_i)} E[u_i(\mathbf{x}_i)] + r(\mathbf{x}_i) \\ \max_{\mathbf{x}_i \in dep(\mathbf{x}^q_i)} E^{yes|\mathbf{x}^q_i,l}\ [u_i(\mathbf{x}_i)] + r(\mathbf{x}_i) \end{cases} \\ &\quad = \max \begin{cases} m \\ \{d_1(\mathbf{x}_i)\ \mu^+(l) + d_2(\mathbf{x}_i) \mid \mathbf{x}_i \in dep(\mathbf{x}^q_i)\}. \end{cases} \end{aligned}$$

where

$$\begin{aligned} &\mu^+(l) = E^{yes|\mathbf{x}^q_i,l}\ [v_i(\mathbf{x}^q_i)], \\ &d_1(\mathbf{x}_i) = (u^\top_i - u^\perp_i) s(\mathbf{x}_i, \mathbf{x}^q_i), \text{ and } d_2(\mathbf{x}_i) = \\ &\quad (u^\top_i - u^\perp_i)(E[\bar{v}_i(\mathbf{x}_i)] - s(\mathbf{x}_i, \mathbf{x}^q_i) E[v_i(\mathbf{x}^q_i)]) + r(\mathbf{x}_i). \end{aligned}$$

$\max_{\mathbf{x}} E^{no|\mathbf{x}^q_i,l}\ [u(\mathbf{x})]$ can be simplified in a similar way.

### 4.2 Mixture of uniforms priors

Specifying prior information over local utility parameters as a mixture of uniform distributions confers several advantages for utility elicitation. With enough components, a mixture of uniforms is flexible enough to approximate many standard distributions; furthermore, it fits nicely with the type of queries we consider here. Because the posterior distribution after a response to a query remains a mixture of uniforms (we only need to update the weights), it is possible to maintain an exact density over utility parameters throughout the elicitation process [2]. Most importantly, we can calculate the optimal query point $l$ analytically. To maximize EPU, we only need to calculate the maximum of

$$\begin{cases} \Pr(yes|l)\ m + \Pr(no|l)\ (d_1(\mathbf{x}_i)\ \mu^-(l) + d_2(\mathbf{x}_i)) \\ \Pr(yes|l)\ (d_1(\mathbf{x}_i)\ \mu^+(l) + d_2(\mathbf{x}_i)) + \Pr(no|l)\ m \end{cases}$$

for each $\mathbf{x}_i \in$ *dep*$(\mathbf{x}^q_i)$. For a given $\mathbf{x}_i$, this expression is a piecewise quadratic function of $l$. Fig. 2 shows an example of such a function for a density with 5 components. The maximum occurs at $l^* = \frac{m - d_2(\mathbf{x}_i)}{d_1(\mathbf{x}_i)}$, if $l^* \in [0, 1]$.

## 5 Empirical Results

We implemented the myopic elicitation strategy using prior densities specified as mixtures of uniform distributions, and tested it on a realistic car-rental problem [4]. The graphical structure of this problem is sufficient to admit fast (around 1 second) EVOI computation; therefore, our approach can readily support interactive real-time preference elicitation.

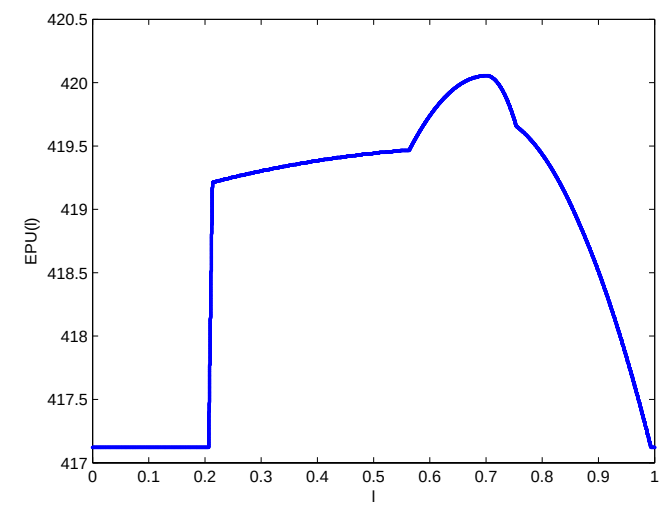

Figure 2: Expected posterior utility as a function of the query point $l$

The car-rental problem is modeled with 26 variables that specify various attributes of a car relevant to typical rental decisions. Variable domains range from 2 to 9 values, resulting in 61,917,364,224 possible configurations. The GAI model consists of 13 local factors, each defined on at most five variables; the model has 378 utility parameters (see [4] for further problem details). Constraints on possible configurations require constraint-based optimization to determine optimal *feasible* configurations of the variables (so even with a precise utility function, optimization is required to determine the best outcome). We use variable elimination to determine best "expected" outcomes.

We experiment with three different types of priors on local utility functions: a (random) mixture of five uniforms, a non-informative uniform density, and a mixture of 10 uniforms which is fitted to approximate a truncated Gaussian distribution with a variance of 0.3 and the mean chosen at random from the interval $[0, 1]$. For each of the three types of priors, we sample 30 different utility functions that are used to generate responses to queries. We then run our elicitation algorithm for 100 queries; for an EVOI query strategy, if the EVOI becomes 0 (which happens after 20-30 queries on average), we choose the next query at random. We compare our myopic EVOI strategy with a "random" query strategy, where a subutility factor and a local query configuration is chosen at random; however, the query point $l$ is set to the expected local utility of the query suboutcome (so $l$ is chosen "intelligently" to give equal odds to either response). Figure 3 summarizes our experimental results for the three different types of priors. All results for EVOI queries are averaged over 30 trials with underlying utility functions sampled from the corresponding priors, while the random strategy results are averaged over 100 trials. Figure 3(b) show (unsurprisingly) that Gaussian priors are quite informative—on average, the initial error (before elicitation) is only slightly greater than 2%, while the uniform priors give an initial error is around 13%. The impact of these differences is normalized in Fig-ure 3(a), which shows how the error decreases as a fraction of the initial error. In all cases, the EVOI strategy is clearly superior to a random query strategy, which at best reduces the error by only 20% after 100 queries. The EVOI strategy cuts the error by at least a half after 50 queries. Though this might seem like a large number of queries, recall that the problem is large (378 utility parameters), and our queries are local comparison queries.

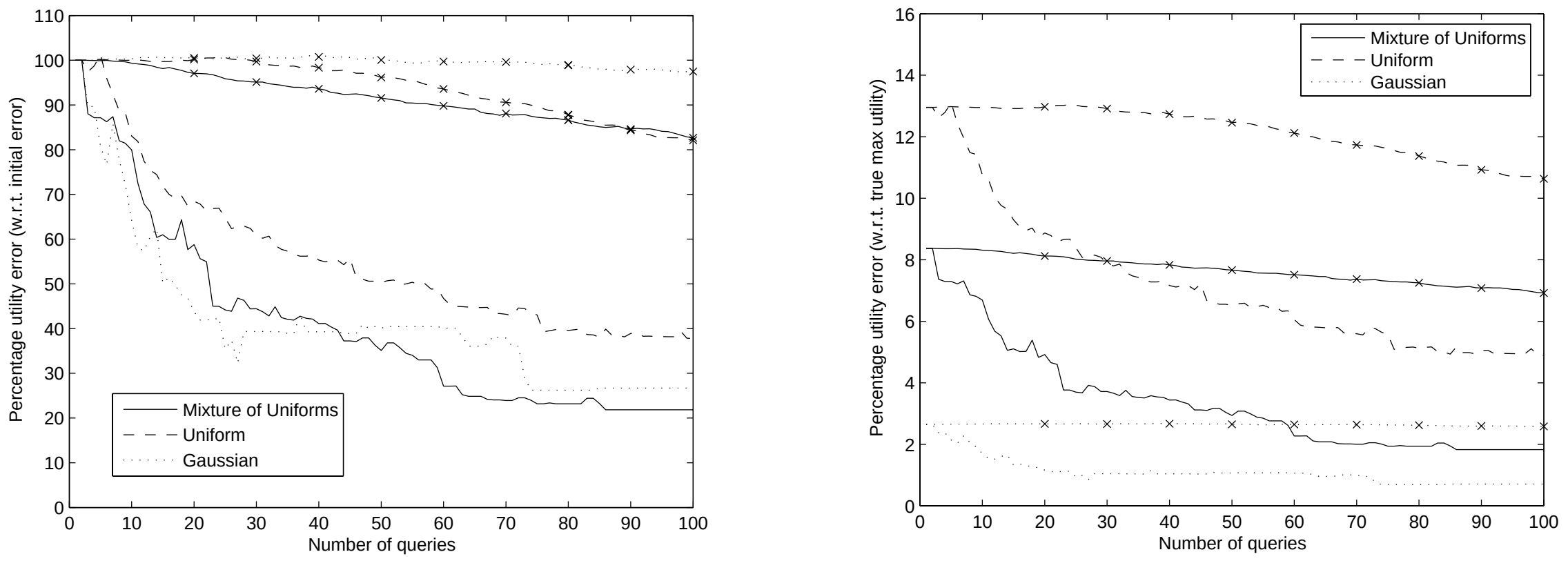

Figure 3: Utility error reduction with queries: (a) as a percentage of the initial error; (b) as a percentage of the optimal utility. EVOI query strategy results are unmarked, and random strategy results are marked with an 'X'.

## 6 Concluding Remarks

We described a new approach to elicitation in GAI models. Unlike previous approaches, we have shown how the graphical structure can be exploited to restrict attention almost exclusively to queries over local outcomes and local standard gambles, thus extending a key advantage of additive models to the generalized case. We have also shown how one can exploit GAI structure to optimize query choice when using myopic EVOI to guide preference elicitation.

A number of directions remains to be explored. Methods for eliciting GAI model structure are of primary importance because a suitable GAI decomposition is a prerequisite for our algorithm [11]. Other directions include incorporating noise models into user responses [2]; developing computationally tractable approximations for computing nonmyopic EVOI in this setting; user case studies and methods for dealing with inconsistency in user responses (though our current method will never ask a query for which a response could be inconsistent); and investigating other decision criteria such as minimax regret [4].

## References

[1] F. Bacchus and A. Grove. Graphical models for preference and utility. *UAI-95*, pp.3–10, Montreal, 1995.

[2] C. Boutilier. A POMDP formulation of preference elicitation problems. *AAAI-02*, pp.239–246, Edmonton, 2002.

[3] C. Boutilier, F. Bacchus, and R. I. Brafman. UCP-Networks: A directed graphical representation of conditional utilities. *UAI-01*, pp.56–64, Seattle, 2001.

[4] C. Boutilier, R. Patrascu, P. Poupart, and D. Schuurmans. Constraint-based optimization with the minimax decision criterion. *CP-2003*, pp.168–182, Kinsale, Ireland, 2003.

[5] U. Chajewska, L. Getoor, J. Norman, and Y. Shahar. Utility elicitation as a classification problem. *UAI-98*, pp.79–88, Madison, 1998.

[6] U. Chajewska, D. Koller, and R. Parr. Making rational decisions using adaptive utility elicitation. *AAAI-00*, pp.363–369, Austin, 2000.

[7] J. S. Dyer. Interactive goal programming. *Management Science*, 19:62–70, 1972.

[8] P. C. Fishburn. Interdependence and additivity in multivariate, unidimensional expected utility theory. *International Economic Review*, 8:335–342, 1967.

[9] P. C. Fishburn. *Utility Theory for Decision Making.* Wiley, New York, 1970.

[10] C. Gonzales and P. Perny. GAI networks for utility elicitation. *KR-04*, 2004.

[11] R. L. Keeney and H. Raiffa. *Decisions with Multiple Objectives: Preferences and Value Trade-offs.* Wiley, NY, 1976.

[12] A. Salo and R. P. Hämäläinen. Preference ratios in multiattribute evaluation (PRIME)–elicitation and decision procedures under incomplete information. *IEEE Trans. on Systems, Man and Cybernetics*, 31(6):533–545, 2001.

[13] O. Toubia, J. Hauser, and D. Simester. Polyhedral methods for adaptive choice-based conjoint analysis. TR-4285-03, Sloan School of Mgmt., MIT, Cambridge, 2003.

[14] J. von Neumann and O. Morgenstern. *Theory of Games and Economic Behavior.* Princeton Univ. Press, Princeton, 1944.

[15] C. C. White, III, A. P. Sage, and S. Dozono. A model of multiattribute decisionmaking and trade-off weight determination under uncertainty. *IEEE Trans. on Systems, Man and Cybernetics*, 14(2):223–229, 1984.